\documentclass[twocolumn,aps,showpacs]{revtex4}
\usepackage{graphicx}
\begin{document}

\title{Absolute Branching Ratio of Beta-Delayed Gamma-Ray Emission of $^{18}N$}
 %\thanks{Supported by USDOE grants No. DE-FG02-91ER40609 and DE-FG02-94ER40870}
\author{R.H. France III}
 \altaffiliation[also at: ]{ A.W. Wright Nuclear Structure Laboratory, P.O. Box 208124
Yale University, 272 Whitney Avenue, New Haven, CT 06520-8124}
\affiliation{Department of Chemistry 
\& Physics, Campus Box 82, Georgia College \& State University,
Milledgeville, Georgia 31061}
\author{Z. Zhao}
\author{M. Gai} 
\altaffiliation[Permanent Address: ]{Department of Physics, 
2152 Hillside Rd. Unit 3046, University of Connecticut, Storrs, CT 06269-3046}
\affiliation{ A.W. Wright Nuclear Structure Laboratory, P.O. Box 208124,
Yale University, 272 Whitney Avenue, New Haven, CT 06520-8124}

\begin{abstract}
The absolute branching ratio of the beta-delayed gamma-ray emission of $^{18}$N was
measured, providing absolute normalization of the  previous work by Olness \textit{et al.}
[J.W.~Olness, \textit{et al.},  
Nuc.~Phys.~{\bf A373}, 13 (1982)] who measured the relative branching ratios for the individual gamma-rays.  We find the
total absolute branching ratio for beta-delayed  gamma emission of $^{18}N$ to be $76.7\pm7.2(stat)
 \pm 5.8(norm)\%$.  A combination of other results suggests a value consistent with our result, but 
 smaller than that calculated by Millener as quoted in Olness \textit{et al.}
\end{abstract}


\maketitle
In their comprehensive study of the beta-decay of $^{18}N$, Olness \textit{et al.} used 
a theoretical calculation for their overall normalization \cite{Olness}.  
While performing a study of the beta-delayed alpha-particle and gamma-ray emissions of 
$^{16}N$ at the National Superconducting Cyclotron Laboratory at Michigan State University 
\cite{tbp}, we were provided with an opportunity to measure the absolute normalization of $^{18}N$
from data taken during a  short calibration run using a  $^{18}N$ beam.  

\begin{figure}
 \includegraphics[height=4in]{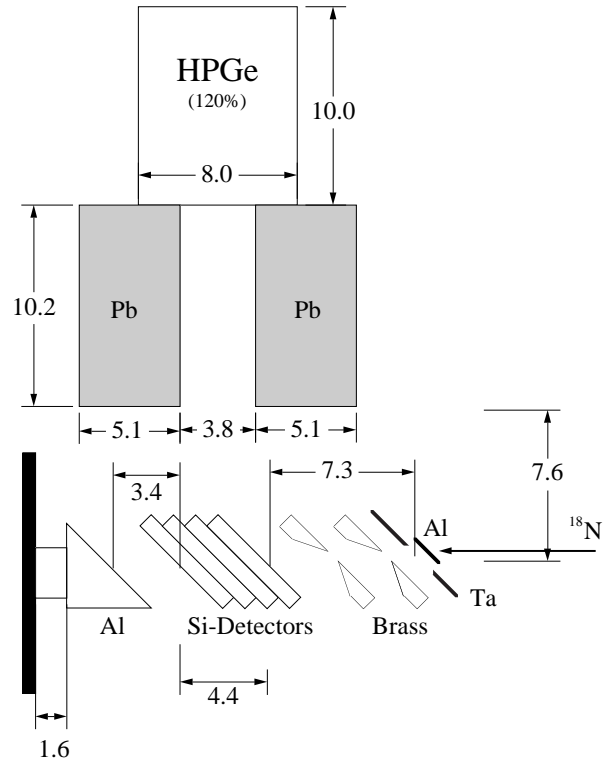}
 \caption{\label{diag}A schematic diagram of the detector setup drawn to scale.  The
Alpha detector stack was enclosed in a vacuum chamber.  Each detector and the two brass collimators
were placed in aluminum frames.  All dimensions shown are in centimeters.}
\end{figure}

Ions of $^{18}N$ were implanted into a stack of four Silicon
surface barrier detectors (SSB), respectively 50 $\mu$m, 25 $\mu$m, 25 $\mu$m, and greater 
than 400 $\mu$m thick, with most of the beam stopping in the three thin detectors.  The
detectors were placed at a $45^\circ$ angle with respect to the
beam, as shown in Fig.~\ref{diag}, to increase their effective thickness
by a factor of $\sqrt{2}$.  Three collimators were placed
upstream; the first was of tantalum, and the second two
were of brass. A 120\% HPGe gamma-ray
detector was placed $17.8\ cm$ away from the center
of the stacked detectors.   The HPGe detector was shielded from the
collimators with $10.2\ cm$ of lead [see Fig.~\ref{diag}], with a $3.8\ cm$ 
diameter hole 
to allow the HPGe detector to see the SSB detectors. In
front of our collimators we placed an aluminum degrader
to slow the ions, permitting them to stop in the detectors.

The secondary beam of $^{18}N$ ions was provided
by the A1200 fragment separator at the National
Superconducting Cyclotron Laboratory at Michigan
State University.  The primary beam was $^{18}O$ at $80\ MeV/u$ 
with a Beryllium target.  The $^{18}N$ fragments with an energy of about $42\ MeV/u$ were separated 
using the A1200 in a momentum-loss achromatic mode providing very good mass separation.

The cyclotron was dephased during every other
second of the run so that data could be taken with 
the beam on and with the beam off.  Different amplifiers were used
during the two phases of the beam cycle, as
the beam particles deposited much more energy in the
detectors than did the alpha particles.

The absolute efficiency of the HPGe detector
was measured by placing  a calibrated $^{65}Zn$ source and a
calibrated $^{228}Th$ source at the position of each alpha
detector,  each collimator, and the aluminum beam stop.
As the detectors were at $45^\circ$ with respect to the beam and the HPGe
detector, the gamma-rays did not pass through the frame of
the detector in which the parent $^{18}N$ nucleus was imbedded.  
During each calibration run, all items in Fig.~\ref{diag} were in place
except the individual detector replaced by the source.  Hence, all absorption
of gamma-rays by the setup is included in the efficiency measurements.
 A calibrated $^{152}Eu$ 
was used to determine the energy dependence of the efficiency.
The HPGe detector was well shielded with lead from the
first two collimators, and the beam, itself, was stopped before
the aluminum beam stop.  

\begin{figure}
 \includegraphics[height=2in]{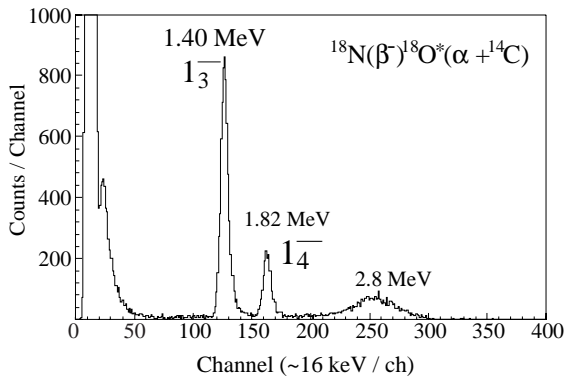}
  \caption{\label{alpha}Measured Beta-Delayed Alpha-Particle Spectrum for $^{18}$N.}
  \end{figure}

\begin{figure}
 \includegraphics[height=2in]{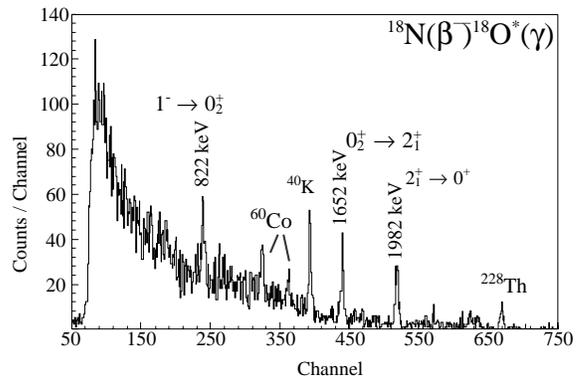}
  \ \caption{\label{gamma}Measured Beta-Delayed Gamma-Ray Spectrum for $^{18}$N.}
  \end{figure}

The yield for alpha particles in the $1.40\ MeV$ peak was measured during the beam-off 
period as shown in Fig.~\ref{alpha}.  From Zhao {\it et al.}\ 
\cite{Zhao} the absolute branching ratio for this decay
is known to be $6.8\pm0.5\%$, which was used for normalization.  From Olness {\it et al.}\ \cite{Olness} the
$1982\ keV$ transition in $^{18}O$ has a relative gamma branching ratio of
$97.95\pm1.96\%$.  Olness {\it et al.}\ \cite{Olness} report a calculated total beta-delayed gamma
emission branching ratio of $85\pm6\%$ leading to an absolute branching ratio of the $1982\ keV$ line
of $84\pm6\%$.  In this work, the absolute branching ratio for the $1982\ keV$ line was measured from
its yield relative to the known branching ratios of the beta-delayed alpha-particle decay
\cite{tbp} and the absolute gamma-ray detection efficiency at $1982\ keV$ [see Fig.~\ref{gamma}].

\begin{figure}
 \includegraphics[height=3in]{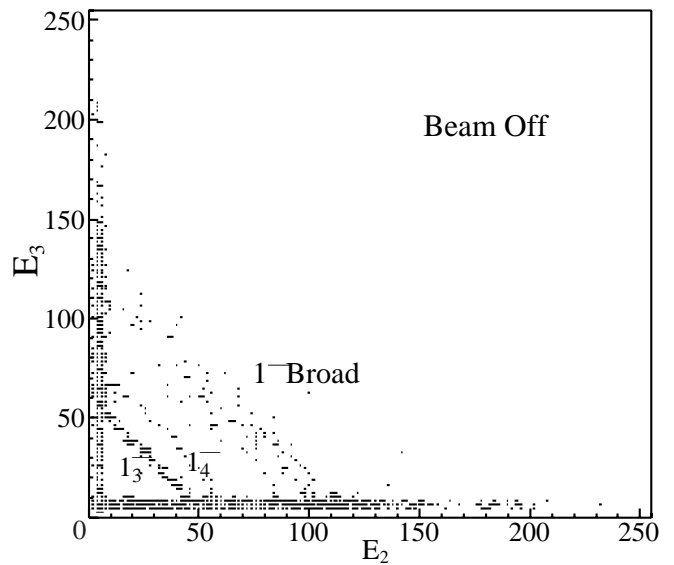}
 \caption{\label{beamoff}Energy detected 
 in detector 2 ($E_2$) vs energy detected in detector 3 ($E_3$) with
beam off.}
\end{figure}

The data were corrected for the
incomplete stopping of alpha particles within the
surface barrier detectors.  We measured this effect by
plotting the energy in detector 1 versus the energy
deposited in detector 2 (and similarly for detectors 2 and
3) as shown in Fig.~\ref{beamoff}.  Here we summed the
events lying within the band $E_2 + E_3 = 1.40\ MeV$
labelled  $1^-_3$ in Fig.~\ref{beamoff}.  This correction was found to be approximately 2\% .  The
detection of gamma-rays from $^{18}N$ ions stopped on the collimators or the aluminum beam
dump was estimated to be in the range of 1\% at $2.6\ Mev$
to unmeasurable at $<1\ MeV$.  We included this
uncertainty in the uncertainty of the HPGe efficiency, 
which is conservatively estimated at 3.3\%.  

The area of the $1982\ keV$ gamma-ray line, calculated using the LEONE program with a gaussian fit was $118.9 \pm 10.9$.  
The absolute efficiency of the HPGe detector was measured to be $8.16 \pm 0.18 \times 10^{-4}$; thus the total number of
$1982\ keV$ gamma-rays emitted during the experiment was $1.46\pm 0.14 \times 10^5$. The total number of alpha-particles,
taking into account all uncertainties listed above, was $1.32 \pm 0.01 \times 10^4$.  Combining these with the known
branching ratios listed above yields a branching ratio for the $1982\ keV$ line of $75.3 \pm 7.1(stat) \pm 5.5(norm)\%$,
with the normalization uncertainty coming from the uncertainty in the alpha-particle branching ratio.  The final 
branching ratio for all gamma-ray emission is thus $76.7 \pm 7.2(stat) \pm 5.8(norm)\%$.  

The allowed beta-delayed emissions of
$^{18}N$ are beta-delayed alpha-particle, gamma-ray, and 
neutron emission. Beta-decay to the ground state of $^{18}O$
is also predicted to be $2.6\%$ \cite{Olness}.  The total branching ratios for alpha-particle emission are
given by Zhao {\it et al.}\ \cite{Zhao} as $12.2 \pm 0.6\%$.  The branching ratio for higher energy neutron
emission is given by Scheller {\it et al.}\ \cite{Sch} as 2\%, and the total branching ratio for neutron
emission is given by Reeder {\it et al.}\ to be $14.3 \pm 2.0\%$\ \cite{Reeder}.  These combine to a total
beta-delayed gamma-ray emission branching ratio of $70.9 \pm 2.1\%$.  This value is consistent with our
measurement of $76.7 \pm 7.2(stat) \pm 5.8(norm)\%$, but is smaller than the theoretical value \cite{Olness}
of $85 \pm 6\%$.  

\begin{acknowledgements}
The authors would like to thank K.S. Lai of Yale University, E.L. Wilds of the University of Connecticut,
R.A. Kryger and J.A. Winger of the
National Superconducting Cyclotron Lab of MSU and K.B. Beard of
Hampton University for their aid in performing the research reported here.
Supported by USDOE grants No. DE-FG02-91ER40609 and DE-FG02-94ER40870.
\end{acknowledgements}

\end{document}